\documentclass[12pt]{article}
\usepackage{epsfig}
\newcommand{\be}{\begin{equation}}
\newcommand{\ee}{\end{equation}}
\newcommand{\bqa}{\begin{eqnarray}}
\newcommand{\eqa}{\end{eqnarray}}
\newcommand{\pslash}{\slash\hspace{-0.55em}}

\begin{document}

\begin{center}
{\LARGE $\psi(3770)$ and $B$ meson exclusive decay $B \rightarrow \psi(3770) K$ in QCD factorization}\\[0.8cm]
{\large Ying-Jia Gao$~^{(a)}$, Ce Meng$~^{(a)}$
and~Kuang-Ta Chao$~^{(b,a)}$}\\[0.5cm]
{\footnotesize (a)~Department of Physics, Peking University,
 Beijing 100871, People's Republic of China}

{\footnotesize (b)~China Center of Advanced Science and Technology
(World Laboratory), Beijing 100080, People's Republic of China}
\end{center}

\vspace{0.5cm}
\begin{abstract}
Belle has observed surprisingly copious production of $\psi(3770)$
in $B$ meson decay $B\to \psi(3770)K$, of which the rate is
comparable to that of $B\to \psi(3686)K$. We study this puzzling
process in the QCD factorization approach with the effect of S-D
mixing considered. We find that the soft scattering effects in the
spectator interactions play an essential role. With a proper
parametrization for the higher twist soft end-point singularities
associated with kaon, and with the S-D mixing angle
$\theta=-12^{\circ}$, the calculated decay rates can be close to
the data. Implications of these soft spectator effects to other
charmonium production in $B$ exclusive decays are also emphasized.

\vspace{0.5cm}
\end{abstract}

\section{Introduction}
The $\psi(3770)$ is the lowest lying charmonium state above the
open charm $D\bar{D}$ threshold. It is expected to be
predominantly the $1^3D_1$ charmonium state with a small admixture
of the $2^3S_1$ component. The $\psi(3770)$ is of great interest
in recent studies of charmonium physics. There are a number of new
measurements and related theoretical issues about the
$\psi(3770)$, e.g. the non $D-\bar D$ decays including charmonium
transitions and decays to light hadrons\cite{bes, cleo2, cleo1}
(see also \cite{kuang}), the radiative transitions to the P-wave
charmonia\cite{cleo3}, the $S-D$ mixing, and the discussions about
the well known $\rho\pi$ puzzle in $J/\psi$ and $\psi(3686)$
decays (see, e.g., \cite{rosner, yuan}).

In this paper, we will focus on another interesting issue about
the $\psi(3770)$. That is the $\psi(3770)$ production in the $B$
meson exclusive decay $B\to \psi(3770)K$, of which the rate is
found by Belle to be surprisingly large\cite{belled}, even
comparable to that of $B\to \psi(3686)K$, and it might seemingly
indicate that this result suggests a large amount of S-D mixing in
the $\psi(3770)$\cite{belled}. But this apparently needs a careful
examination.

It is generally believed that if the virtual charmed meson pair
components are neglected the two states $\psi(3686)$ and
$\psi(3770)$ can be approximately expressed as
\bqa\mid\psi'\rangle \equiv \mid \psi(3686) \rangle &=& cos\theta
\mid c\bar{c}(2^3S_1)\rangle + sin\theta
\mid c\bar{c}(1^3D_1)\rangle, ~~~~~\nonumber    \\
\mid \psi''\rangle \equiv\mid\psi(3770) \rangle &=& cos\theta \mid
c\bar{c}(1^3D_1)\rangle - sin\theta \mid c\bar{c}(2^3S_1)\rangle.
\eqa
The S-D mixing angle has been estimated by using the ratio of the
leptonic decay widths \cite{pdg} of $\psi(3686)$ and $\psi(3770)$.
Nonrelativistic potential model calculations give two solutions:
$\theta\approx -10^{\circ}$ to $-13^{\circ}$ or $\theta\approx
+30^{\circ}$ to $+26^{\circ}$ \cite{rosner,kuang,ding}. The small
mixing angle is compatible with the results obtained in models
with coupled channel effects \cite{eichten, heikkila} and is
favored by the E1 transition $\psi'\rightarrow \gamma\chi_{cJ}$
data also\cite{ding}.

Belle Collaboration \cite{belled} has observed $\psi(3770)$ in the B
meson decay $B^+\rightarrow\psi(3770)K^+$ with a branching ratio,
\bqa {\mathrm{Br}} (B^+ \rightarrow \psi^{''} K^+) &=& (0.48\pm
0.11\pm 0.07) \times 10^{-3},\label{psi3770}\eqa
which is comparable to that of $\psi(3686)$~\cite{pdg},
\bqa {\mathrm{Br}} (B^+ \rightarrow \psi^{'} K^+) &=&(0.66\pm 0.06
)\times 10^{-3}.
  \label{psi3686}\eqa
This is quite surprising, since conventionally the $\psi(3770)$
and $\psi(3686)$ are regarded as predominantly the $1^3D_1$ and
$2^3S_1$ $c\bar c$ states respectively, and the coupling of
$1^3D_1$ to the $c \bar c$ vector current in the weak decay
effective hamiltonian is much weaker than that of $2^3S_1$ in the
naive factorization approach \cite{BSW}. If this experimental
result is really due to a large S-D mixing, as suggested in
\cite{belled}, then it is found in Ref.\cite{liu} that an
unexpectedly large S-D mixing angle $\theta=\pm 40^{\circ}$ would
be required by fitting the observed ratio of $B\to \psi(3770)K$ to
$B\to \psi(3686)K$ decay rates, when the $D$ wave contribution is
neglected. This is in serious contradiction with all other
experimental and theoretical studies, and, in particular, with the
newly measured E1 transition rates for $\psi(3770)\to
\gamma\chi_{cJ} (J=0,1,2)$, for which the CLEO results\cite{cleo3}
are $172\pm 30, 70\pm 17, <21$~KeV respectively for J=0,1,2
whereas the corresponding calculations are $386, 0.32, 66$~KeV for
$\theta=-40^{\circ}$ and $52, 203, 28$~KeV for
$\theta=+40^{\circ}$\cite{liu}. So, based on the naive
factorization, the use of large S-D mixing to explain the Belle
data for $B\to \psi(3770)K$ should be ruled out. The next question
is, can we explain the Belle data by considering the
nonfactorizable contributions to these decay rates?

In the following, we will study this problem in the QCD
factorization approach\cite{BBNS1,BBNS2,BBNS3} including
nonfactorizable contributions. We will first give the decay rate
of $B\to \psi(3770) K$ based on the assumption that $\psi(3770)$
is a pure D-wave charmonium state. Then we take the S-D mixing
into account. Finally, we will consider the higher twist effects.

\section{$B\rightarrow \psi(3770) K$ decay in QCD
factorization}\label{s2}

The effective Hamiltonian for this decay mode is written
as\cite{BBL2}
 \be
H_{\mathrm{eff}} = \frac{G_F}{\sqrt{2}} \Bigl( V_{cb} V_{cs}^* (C_1
{\cal O}_1 +C_2 {\cal O}_2 ) -V_{tb} V_{ts}^* \sum_{i=3}^{10} C_i
{\cal O}_i \Bigr).
 \ee
Here $C_i$'s are the Wilson coefficients which can be evaluated by
the renormalization group approach \cite{BBL2} and the results at
$\mu=4.4$ GeV are listed in Tab. 1. The relevant operators ${\cal
O}_i$ in $H_{\mathrm{eff}}$ are given by
 \bqa
&& {\cal O}_1=(\overline{s}_{\alpha} b_{\beta})_{V-A} \cdot
(\overline{c}_{\beta} c_{\alpha})_{V-A},\qquad\qquad~ {\cal
O}_2=(\overline{s}_{\alpha} b_{\alpha})_{V-A} \cdot
(\overline{c}_{\beta} c_{\beta})_{V-A},
 \nonumber\\
&& {\cal O}_{3(5)}=(\overline{s}_{\alpha} b_{\alpha})_{V-A} \cdot
\sum_q (\overline{q}_{\beta} q_{\beta})_{V-A(V+A)},~ {\cal
O}_{4(6)}=(\overline{s}_{\alpha} b_{\beta})_{V-A} \cdot \sum_q
(\overline{q}_{\beta} q_{\alpha})_{V-A(V+A)},
\\
&& {\cal O}_{7(9)}={3\over 2}(\overline{s}_{\alpha}
b_{\alpha})_{V-A} \cdot \sum_q e_q (\overline{q}_{\beta}
q_{\beta})_{V+A(V-A)},~ {\cal O}_{8(10)}={3\over
2}(\overline{s}_{\alpha} b_{\beta})_{V-A} \cdot \sum_q e_q
(\overline{q}_{\beta} q_{\alpha})_{V+A(V-A)}.\nonumber
 \eqa

\begin{table}[t]
\begin{center}
\begin{tabular}{ c | c c c c c c }
   \hline
    &$C_1$ & $C_2$ & $C_3$ & $C_4$
 & $C_5$ & $C_6$ \\
 \hline
  LO & 1.144 & -0.308& 0.014 & -0.030 & 0.009 & -0.038  \\
  NDR & 1.082 & -0.185 & 0.014 & -0.035 & 0.009 & -0.041 \\
 \hline
 \end{tabular}
\caption{ {Leading-order(LO) and Next-to-leading-order(NLO) Wilson
coefficients in the NDR scheme (See Ref.\cite{BBL2}) with
$\mu=4.4$ GeV and $\Lambda^{(5)}_{\overline{\rm MS}}=225$ MeV.}}
 \label {wilson}
\end{center}
\end{table}

We treat the charmonium as a color-singlet non-relativistic $c\bar
c$ bound state. Let $p_\mu$ be the total 4-momentum of the
charmonium and $2q_\mu$ be the relative 4-momentum between $c$ and
$\bar c$ quarks. For D-wave charmonium, because the wave function
and its first derivative at the origin vanish,
$\mathcal{R}_1(0)\!\!=\!\!0$,
$\mathcal{R}_1^{\prime}(0)\!\!=\!\!0$, which correspond to the
zeroth and the first order in $q$, we must expand the amplitude to
second order in $q$. Thus we have (see, e.g., \cite{kuhn})
 \bqa
 \label{amp}
\mathcal{M}(B\to\!\! {}^3\!D_1(c\bar
c))\!=\frac{1}{2}\sum_{L_z,S_z}\!\langle 2L_z;1S_z|1J_z\rangle
 \!\int\!\!\frac{\mathrm{{d}}^4 q}{(2 \pi)^3}q_\alpha q_\beta \nonumber\\
 \times \delta\!(q^0\!\!-\!\!\frac{|\vec{q}|^2}{M}\!)\psi_{2M}^\ast\!(q)
 \mathrm{Tr}[\mathcal{O}^{\alpha\beta}\!(0)P_{1S_z}\!(p,\!0)
\!+\!\mathcal{O}^\alpha\!(0)P^\beta_{1S_z}\!(p,\!0)\nonumber\\
+\mathcal{O}^\beta\!(0)P^\alpha_{1S_z}\!(p,\!0)+\!\mathcal{O}\!(0)P^{\alpha\beta}_{1S_z}\!(p,\!0)],
 \eqa
where $\mathcal{O}(q)$ represents the rest of the decay matrix
element. The spin-triplet projection operators $P_{1S_z}(p,q)$ is
constructed in terms of quark and anti-quark spinors as
 \bqa P_{1S_z}(p,q)\!=\!\!\sqrt{\!\frac{3}{m}}\!\sum_{s_1,s_2}\!\!v(\frac{p}{2}\!-\!q,\!s_2)
 \bar u(\!\frac{p}{2}\!+\!q,\!s_1) \!\langle s_1;\!s_2|1S_z\!\rangle,
 \eqa
 and
 \bqa
\mathcal{O}^\alpha(0)\!&=&\!\frac{\partial\mathcal{O}(q)}{\partial
q_\alpha}|_{q=0},~~~~~~
\mathcal{O}^{\alpha\beta}(0)\!=\!\frac{\partial^2
\mathcal{O}(q)}{\partial q_\alpha\partial q_\beta}|_{q=0},\nonumber\\
 P^\alpha_{1S_z}(p,0)\!&=&\!\frac{\partial P_{1S_z}(p,q)}{\partial q_\alpha}
 |_{q=0},~~
 P^{\alpha\beta}_{1S_z}(p,0)\!=\!\frac{\partial^2 P_{1S_z}(p,q)}{\partial q_\alpha\partial q_\beta}
 |_{q=0}\,.
  \eqa

After $q^0$ is integrated out, the integral in Eq.(\ref{amp}) is
proportional to the second derivative of the D-wave wave function
at the origin by
 \bqa
\int\!\frac{\mathrm{{d}}^3 q}{(2 \pi)^3}q^\alpha q^\beta
\psi_{2m}^\ast(q) =e^{\ast\alpha\beta}_m\sqrt{\frac{15}{8\pi}}
\mathcal{R}^{''}_D(0), \eqa where $e^{\alpha\beta}_m$ is the
polarization tensor of an angular momentum-2 system and the value of
$\mathcal{R}^{''}_D(0)$ for charmonia can be found in, e.g.,
Ref.\cite{quig}.

The spin projection operators $\!P_{1S_z}(p,0)\!$ ,
$\!P^\alpha_{1S_z}(p,0)\!$ and $\!P^{\alpha\beta}_{1S_z}(p,0)\!$ can
be written as\cite{kuhn}
 \bqa
P_{1S_z}(p,0)&\!\!=\!\!&\sqrt{\frac{3}{4 M}}\pslash
\varepsilon^\ast(S_z)(\pslash p+M),\\
P^\alpha_{1S_z}(p,0)&\!\!=\!\!\!&\sqrt{\frac{3}{4 M^3}}[\pslash
\varepsilon^\ast(S_z)(\pslash
p+\!M)\gamma^\alpha\!+\!\gamma^\alpha \pslash
\varepsilon^\ast(S_z)(\pslash p+\!M)],\\
P^{\alpha\beta}_{1S_z}(p,0)&\!\!=\!\!\!&\sqrt{\frac{3}{4
M^5}}[\gamma^\beta\pslash \varepsilon^\ast(S_z)(\pslash
p+\!M)\gamma^\alpha\!+\!\gamma^\alpha \pslash
\varepsilon^\ast(S_z)(\pslash p+\!M) \gamma^\beta],\nonumber
 \eqa
where we have made use of the non-relativistic approximation for
the charmonium mass $M\simeq 2 m$. Here $m$ is the charmed quark
mass.

As for the light meson kaon, we describe it relativistically by
light-cone distribution amplitudes (LCDAs) \cite{BBNS3} up to
twist-3 level:
\bqa
   \langle K(p)|\bar s_\beta(z_2)\,d_\alpha(z_1)|0\rangle&=&\frac{i f_K}{4} \int_0^1dxe^{i(y\,p\cdot z_2+\bar y \,p\cdot z_1)}\nonumber\\
   &&\hspace*{-3.5cm}\times\Bigl\{ \pslash{p}\,\gamma_5\,\phi_K(y)- \mu_K\gamma_5 ( \phi_K^p(y) - \sigma_{\mu\nu}\,p^\mu
(z_2-z_1)^\nu\,
    \frac{\phi_K^\sigma(y)}{6} ) \Bigr\}_{\alpha\beta},
\label{kaon}
  \eqa
where $y$ and $\bar{y}=1-y$ are the momentum fractions of the $s$
and $\bar{d}$ quarks inside the K meson respectively. Here the
chirally enhanced mass scale $\mu_K = {m_K}^2/(m_s(\mu)+m_d(\mu))$
is comparable to $m_b$, which ensures that the twist-3 spectator
interactions are numerically large, though they are suppressed by
$1/m_b$. The twist-2 LCDA $\phi_K(y)$ and the twist-3 ones
$\phi_K^p(y)$ and $\phi_K^\sigma(y)$ are symmetric under
$y\leftrightarrow\bar{y}$ in the limit of SU(3) isospin symmetry. In
practice, we choose the asymptotic forms for these LCDAs,
 \bqa \label{asy}
 \phi_K (y)=\phi_K^\sigma(y)=6y(1-y),\hspace{0.6cm}\phi_K^p (y)=1.
 \eqa

In the naive factorization, we neglect the strong interaction
corrections and the power corrections in
$\Lambda_{\mathrm{QCD}}/m_b$. Then the decay amplitude can be
written as \bqa
 i\mathcal{M}_0= - f_D m_{\psi^{''}} (2p_B\cdot\varepsilon^\ast) F_1 (m_{\psi^{''}}^2) \frac{G_F}{\sqrt{2}}
 \Bigl[ V_{cb} V_{cs}^* (C_2
+\frac{C_1}{N_c} ) -V_{tb} V_{ts}^* (C_3 + \frac{C_4}{N_c} +C_5
+\frac{C_6}{N_c}) \Bigr], \label{tree}
 \eqa
where $N_c$ is the number of colors. We do not include the effects
of the electroweak penguin operators since they are numerically
small. The form factors for $B \rightarrow K$ are given as
  \bqa
\langle K(p_K) | \overline{s} \gamma_{\mu} b| B(p_B)\rangle=\Bigl[
(p_B +p_K)_{\mu} -\frac{m_B^2-m_K^2}{p^2} p_{\mu} \Bigr] F_1 (p^2)
+ \frac{m_B^2-m_K^2}{p^2} p_{\mu} F_0 (p^2),
 \label{vmu2}
  \eqa
where $p= p_B -p_K$ is the momentum of $\psi^{''}$ with $p^2 =
m_{\psi^{''}}^2$. The kaon mass will be neglected in the heavy
quark limit and we will use the approximate relation
${F_0(m_{\psi^{''}}^2)}/{F_1 (m_{\psi^{''}}^2)}=\! 1-r$
\cite{chay, cheng}, where $r\!=\!m_{\psi^{''}}^2/m_B^2$, to
simplify the amplitude in our calculations.

\begin{figure}[t]
\begin{center}
\vspace{-3.5cm}
\includegraphics[width=14cm,height=18cm]{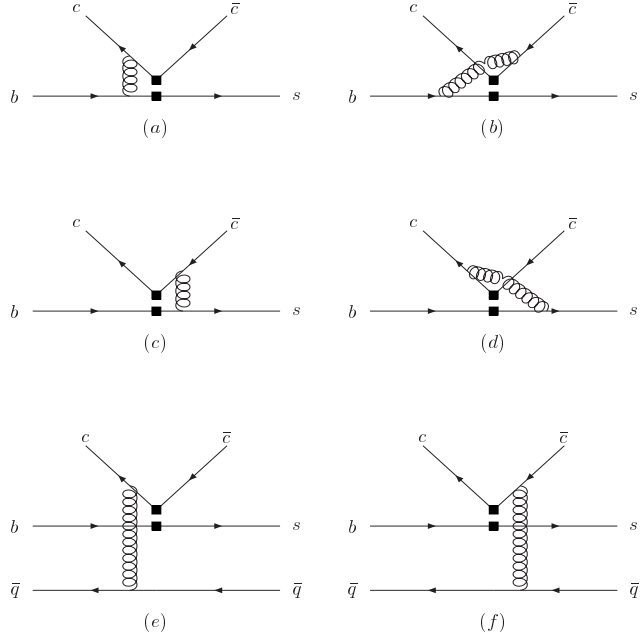}
\vspace{-5.5cm}
\end{center}
\caption{ Feynman diagrams for nonfactorizable corrections to $B
\rightarrow \psi^{''} K$ decay.} \label{fvs}
\end{figure}

As we can easily see in Eq.~(\ref{tree}), this amplitude is
unphysical because the Wilson coefficients depend on the
renormalization scale $\mu$ while the decay constant and the form
factors are independent of $\mu$. This is the well known problem
with the naive factorization. However, if we include the order
$\alpha_s$ corrections, it turns out that the $\mu$ dependence of
the Wilson coefficients is largely cancelled and the overall
amplitude is insensitive to the renormalization scale. Taking the
nonfactorizable order $\alpha_s$ strong-interaction corrections in
Fig.~\ref{fvs} into account, the full decay amplitude for $ B
\rightarrow \psi^{''} K$ within the QCD factorization approach is
written as
 \be
  i\mathcal{M} = f_D m_{\psi^{''}} (2p_B\cdot\varepsilon^\ast) F_1 (m_{\psi^{''}}^2) \frac{G_F}{\sqrt{2}}
 \Bigl[ V_{cb} V_{cs}^* a_2
 -V_{tb} V_{ts}^* (a_3 + a_5) \Bigr], \label{full}
 \ee
where the coefficients $a_i$ ($i=2,3,5$) in the naive dimension
regularization (NDR) scheme are given by
\bqa
 && a_2=-(C_2 +\frac{C_1}{N_c}) +\frac{\alpha_s}{4\pi}
\frac{C_F}{N_c} C_1 \Bigl( -12\ln \frac{m_b}{\mu}+2 + f_I + f_{II}
\Bigr), \nonumber \\
&&a_3=-(C_3 +\frac{C_4}{N_c}) +\frac{\alpha_s}{4\pi}
\frac{C_F}{N_c} C_4 \Bigl( -12 \ln \frac{m_b}{\mu}+2 +f_I + f_{II}
\Bigr),   \label{ai}\\
&&a_5=-(C_5 +\frac{C_6}{N_c}) -\frac{\alpha_s}{4\pi}
\frac{C_F}{N_c} C_6 \Bigl( -12 \ln \frac{m_b}{\mu}-10 +f_I +
f_{II} \Bigr).\nonumber
  \eqa

The function  $f_I$ in Eq.(\ref{ai}) is calculated from the four
vertex diagrams (a,b,c,d) in Fig.~\ref{fvs},
 \bqa
  f_I&=&\int_0^1 dx\int_0^{1-x} dy\  \Biggl[-6\ln \bigl[(x+\frac{y}{2})(x+\frac{r y}{2})\frac{y}{2}
  \bigl((r-1)x+\frac{r y}{2}\bigr)  \bigr]\nonumber\\
  &&-\frac{3}{5}(1-r)^2 x^2 y^2(\frac{1}{(x+\frac{y}{2})^2(x+\frac{r
  y}{2})^2}
  +\frac{1}{(\frac{y}{2}((r-1)x+\frac{r y}{2}))^2})\nonumber\\
  &&-2ry(1-y)\bigl(\frac{1}{(x+\frac{y}{2})(x+\frac{r
  y}{2})}+\frac{1}{\frac{y}{2}((r-1)x+\frac{r y}{2})} \bigr)\nonumber\\
  &&-2\bigl(\frac{(1+r)x-r(2-y)}{x+\frac{r y}{2}}+\frac{(r-1)x-r(2-y)}{(r-1)x+\frac{r y}{2}}\bigr)\nonumber\\
  &&-2r(1-r)x y^2\bigl(\frac{1}{(x+\frac{y}{2})(x+\frac{r
  y}{2})^2}-\frac{1}{\frac{y}{2}((r-1)x+\frac{r y}{2})^2}\bigr) \nonumber\\
  &&+\frac{2}{5}r(1-r)^2x^2y^2\bigl(\frac{1}{(x+\frac{y}{2})(x+\frac{r
  y}{2})^3}+\frac{1}{\frac{y}{2}((r-1)x+\frac{r y}{2})^3}\bigr) \Biggr],
 \eqa
where $r=m_{\psi^{''}}^2/m_B^2$.

The function $f_{II}$ in Eq.(\ref{ai}) is calculated from the two
spectator interaction diagrams (e,f) in Fig.~\ref{fvs} and it is
given by
 \bqa
  f_{II}&=& \frac{16\pi^2}{N_c} \frac{f_K f_B}{m_B^2 F_1
(m^2_{\psi^{''}})} \int_0^1 d\xi \frac{\phi_B (\xi)}{\xi} \int_0^1
dy \phi_K(y)\nonumber\\
&&\times\Bigl[-\frac{1}{10}\frac{1}{(1-r)(1-y)}-\frac{r}{(1-y)^2(1-r)^2}\Bigr],
\label{fII}
 \eqa
 where $\phi_B$ is the light-cone wave functions for the $B$ meson.
 The spectator contribution depends on the wave function
$\phi_B$ through the integral \be
 \int_0^1 d\xi \frac{\phi_B (\xi)}{\xi} \equiv
\frac{m_B}{\lambda_B}.
 \ee
Since $\phi_B (\xi)$ is appreciable only for $\xi$ of order
$\Lambda_{\mathrm{QCD}}/m_B$, $\lambda_B$ is of order
$\Lambda_{\mathrm{QCD}}$. We will follow Ref.~\cite {BBNS3} to
choose $\lambda_B\approx 300$ MeV in the numerical calculation.

It is easily seen from (\ref{fII}) that there is logarithmic end
point singularity in the integration over $y$ when $y\rightarrow
1$. It breaks down the factorization even at the leading twist
level. It implicates that the soft mechanisms may be important to
this decay mode. To estimate these soft effects, we simply
parameterize the end point singularity as
\bqa X\equiv\int^1_0 {dy\over y}={\rm
ln}\left({m_B\over\Lambda_h}\right), \label{logdiv} \eqa
where $\Lambda_h\sim 500$ MeV is the typical momentum scale
associated with the light quark in the B meson. Furthermore, since
the virtuality of the gluon exchanged between the spectator quark
and the charm (or anti-charm) quark is $\Lambda_h m_b$,  we should
multiply a factor $\alpha_s(\sqrt{\Lambda_h m_b})C_i(\sqrt{\Lambda_h
m_b})/(\alpha_s(\mu)C_i(\mu))$ to $f_{II}$ in Eq(\ref{ai}), where
$\mu\sim m_b$ is the scale at which we evaluate those vertex
corrections.

The decay constant $f_D$ is calculated through the potential models,
 \bqa
 f_D=\frac{10\sqrt{3}}{\sqrt{2\pi m_{\psi^{''}}}}\frac{R_D^{''}(0)}{m_{\psi^{''}}^2}.
\eqa
 For numerical analysis, we choose $F_1
(m_{\psi^{''}}^2) = 0.97$\cite {ball} and use the following input
parameters:
 \bqa
&&m_b=4.8 \ \mbox{GeV}, \ \ m_B=5.28 \ \mbox{GeV}, \ \
m_{\psi^{''}}=3.77 \ \mbox{GeV}, \nonumber \\
&&f_B = 216 \ \mbox{MeV}\cite{gray}, \ \ f_K = 160 \ \mbox{MeV}.
 \eqa

Then we get the branching ratio: ${\mathrm{Br}} (B \rightarrow
\psi^{''} K) = 1.13 \times 10^{-5}$. The theoretical calculation
is about 40 times lower than the experimental data
(\ref{psi3770}).

\section{$B\rightarrow \psi^{\prime} K$ decay }\label{s3}

The calculation of the branching ratio for $B\rightarrow
\psi^{\prime} K$ decay is similar to that for $B\rightarrow
\psi^{\prime\prime} K$. If one treats $\psi^{\prime}$ as a pure
2$S$-state, the only modification needed to do is to expand the
decay amplitudes to zeroth order in the $q$-expansion. Thus the
amplitudes will be proportional to the S-wave wave function at the
origin through the integration
\bqa \int\!\frac{\mathrm{{d}}^3 q}{(2 \pi)^3}\psi_{2S}^\ast(q)
=\sqrt{\frac{1}{4\pi}} \mathcal{R}_{2S}(0).\eqa

The full decay amplitude for $ B \rightarrow \psi^{'} K$ within the
QCD factorization approach is written as
 \be
  i\mathcal{M}^{'} = \sqrt{\frac{3}{\pi
m_{\psi^{'}}}}R_{2S}(0) m_{\psi^{'}} (2p_B\cdot\varepsilon^\ast) F_1
(m_{\psi^{'}}^2) \frac{G_F}{\sqrt{2}}
 \Bigl[ V_{cb} V_{cs}^* a_2^{'}
 -V_{tb} V_{ts}^* (a_3^{'} + a_5^{'}) \Bigr], \label{full}
 \ee
where the coefficients $a_i^{'}$ ($i=2,3,5$) in the naive
dimension regularization(NDR) scheme are given by
  \bqa
 && a_2^{'}=(C_2 +\frac{C_1}{N_c}) +\frac{\alpha_s}{4\pi}
\frac{C_F}{N_c} C_1 \Bigl( 12\ln \frac{m_b}{\mu}-2 + f_I^{'} +
f_{II}^{'}
\Bigr), \nonumber \\
&&a_3^{'}=(C_3 +\frac{C_4}{N_c}) +\frac{\alpha_s}{4\pi}
\frac{C_F}{N_c} C_4 \Bigl( 12\ln \frac{m_b}{\mu}-2 + f_I^{'} +
f_{II}^{'}
\Bigr),   \label{ai3686}\\
&&a_5^{'}=(C_5 +\frac{C_6}{N_c}) -\frac{\alpha_s}{4\pi}
\frac{C_F}{N_c} C_6 \Bigl( 12\ln \frac{m_b}{\mu}+10 + f_I^{'} +
f_{II}^{'} \Bigr).\nonumber
  \eqa
Again, the vertex corrections associated with $F_I$ are evaluated
at renormalization scale $\mu\approx m_b$ and the spectator
interactions associated with $F_{II}$ are evaluated at
$\sqrt{\Lambda_h m_b}$.

The function  $f_I^{'}$ and $f_{II}^{'}$ in Eq.(\ref{ai3686}) have
following forms,
 \bqa
   f_I^{'}&=&\int_0^1 dx\int_0^{1-x} dy\  \Bigl[6\ln \bigl[(x+\frac{y}{2})(x+\frac{z y}{2})\frac{y}{2}
  \bigl((z-1)x+\frac{z y}{2}\bigr)  \bigr]\nonumber\\
  &+&{   } 4- \frac{2x(1-z)}{x+\frac{z y}{2}}
   +\frac{z y-(1-z)x}{\frac{1}{2}((z-1)x+\frac{z y}{2})}
   \Bigr],\nonumber\\
     f_{II}^{'}&=& \frac{8\pi^2}{N_c} \frac{f_K f_B}{m_B^2 F_1
(m_{\psi^{'}}^2)(1-z)} \int_0^1 d\xi \frac{\phi_B (\xi)}{\xi}
\int_0^1 dy \frac{\phi_K(y)}{1-y}, \label{fII'}
  \eqa
where $z=m_{\psi^{'}}^2/m_B^2$ and $F_1 (m_{\psi^{'}}^2)=0.91$.
One can easily get the functions in (\ref{fII'}) by using the
known results of $B \rightarrow J/\psi K$ in Ref.~\cite{chay},
where $J/\psi$ is described by LCDAs. We only need to replace the
decay constant $f_{J/\psi}$ by $f_{2S}=\sqrt{\frac{3}{\pi
m_{\psi^{'}}}}R_{2S}(0)$ and choose the non-relativistic limit
form $\phi_{NR}(u)=\delta(u-1/2)$ for LCDAs of $J/\psi$ as in
Ref.~\cite{chay}.

According to Eq. (1) we can write down the ratio of the decay
rates directly:
 \bqa
  R=\frac{Br(B\rightarrow
  \psi^{''}K)}{Br(B\rightarrow\psi^{'}K)}&=&(\frac{1-r}{1-z})\mid\frac{-i\mathcal{M}^{'}\times sin\theta
+i\mathcal{M}\times cos\theta}{i\mathcal{M}^{'}\times cos\theta
+i\mathcal{M}\times sin\theta }\mid^2.
 \eqa

The ratio determined by experimental data is $R\approx0.72$.
Comparing it with our calculation, we can find the mixing angle is:
$\theta=-26^\circ$ or $\theta=+59^\circ$. But the absolute branching
ratio of $B\rightarrow \psi^{''}K$ is $5.9 \times 10^{-5}$, which is
still about one order of magnitude lower than experimental data in
Eq.~(\ref{psi3770}).

\section{Higher twist effects and end point singularities}

In last two sections, we have only considered the leading twist
spectator interactions. Generally, the contributions arising from
higher twist LCDAs of K meson will be suppressed by powers of
$1/m_b$. However, as we have mentioned, the chirally enhanced
scale $\mu_K\sim m_b$ in (\ref{kaon}) ensures that the twist-3
contributions may be numerically large. It was discussed several
years ago that these contributions may play important roles in the
process of B meson to S-wave charmonia decays\cite{cheng}. Here we
consider the higher twist effects in D-wave charmonium production
as well.

The distribution amplitude of kaon to twist-3 have been given in
(\ref{kaon}), then we can find the twist-3 modifications to
$f_{II}$ and $f^{'}_{II}$ to be
 \bqa
  f^3_{II}&=& -\frac{16\pi^2}{N_c} \frac{f_K f_B}{m_B^2 F_1
(m_{\psi^{''}}^2)}\frac{r_K}{(1-r)^2} \int_0^1 d\xi \frac{\phi_B
(\xi)}{\xi} \int_0^1 dy \frac{\phi_K^\sigma(y)}{6} (\frac{r}{(r-1)\
y^3}+\frac{1}{10\ y^2}),\nonumber\\
f^{3'}_{II}&=& \frac{8\pi^2}{N_c} \frac{f_K f_B}{m_B^2 F_1
(m_{\psi^{'}}^2)}\frac{r_K}{(1-z)^2} \int_0^1 d\xi \frac{\phi_B
(\xi)}{\xi} \int_0^1 dy \frac{\phi_K^\sigma(y)}{6}\frac{1}{\
y^2}\,,
 \eqa
where $r_K=2\mu_K/m_b$. Here, we can see there exist logarithmic
end point singularities in both $f^3_{II}$ and $f^{3'}_{II}$. More
seriously, there emerges linear singularity in function $f^3_{II}$
and we will parameterize it just like what we have done for the
logarithmic ones:
\bqa
\int_0^1\frac{dy}{y^2}=\frac{m_B}{\Lambda_h}.\label{lineardiv}\eqa

\begin{figure}[t]
\begin{center}
\includegraphics[width=13cm]{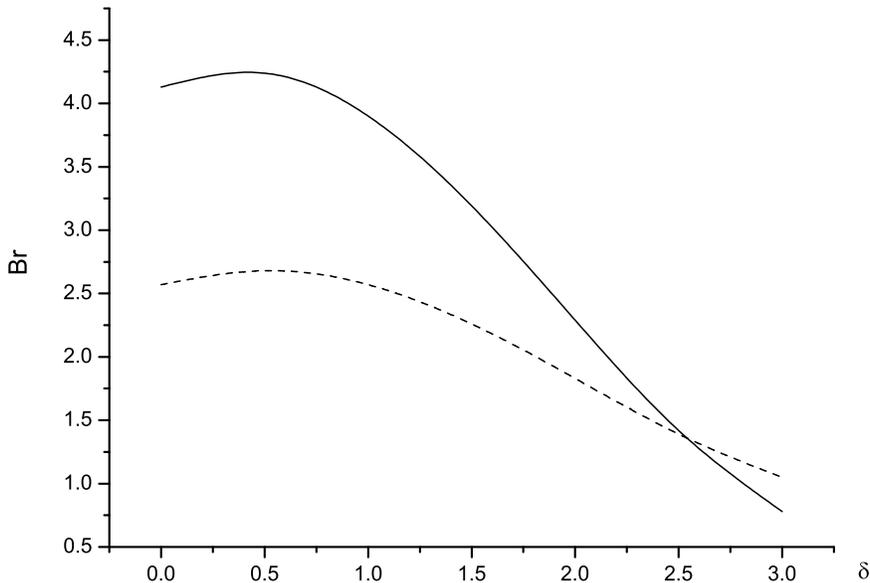}
\end{center}
\caption{Branching ratios of $B\to \psi(3770) K$ and $B\to
\psi(3686) K$ (in units of $10^{-4}$) as functions of $\delta$.
The dashed line is for $B\to \psi(3770) K$ and the solid line for
$B\to \psi(3686) K$.} \label{fig36}
\end{figure}

It is implicit that these singularities can be regularize by the
gluon or light quark offshellness of order $\Lambda_h^2$ when we
use (\ref{logdiv}) and (\ref{lineardiv}). So when the offshellness
is negative, the logarithmic singularity will receive large
complex contributions from the implicit pole in the region of
integration. They are common effects in soft rescattering
processes. Then following \cite{bbns}, we rewrite (\ref{logdiv})
as
\bqa
X\equiv\int^1_0 {dy\over y}={\rm
ln}\left({m_B\over\Lambda_h}\right)+t, \label{logdiv2} \eqa
where $t$ is a complex free parameter and we choose $|t|$ varying
from 3 to 6 as suggested in \cite{cheng}. Setting $|t|=4.5$,
$0\leq\delta\leq \pi$, and the S-D mixing angle
$\theta=-12^{\circ}$, we can get the branching ratio curves of
$B\to \psi(3770) K$ and $B\to \psi(3686) K$, which are shown in
fig.\ref{fig36}.

From fig.\ref{fig36} we see that in the region with small $\delta$
the branching ratios are not very sensitive to the value of
$\delta$. With a value of, say $\delta =\pi /8$, the branching
ratios of $B\to \psi(3770) K$ and $B\to \psi(3686) K$ are found to
be: \bqa {\mathrm{Br}} (B^+ \rightarrow
\psi^{''} K^+) &=&  2.68\times 10^{-4},\nonumber\\
 {\mathrm{Br}} (B^+ \rightarrow \psi^{'}
K^+) &=&  4.25\times 10^{-4}.
  \eqa
From these values we can get $R=0.63$, which fits the experimental
data quite well. At the same time, the absolute branching ratios are
both close to the experimental data. So we may conclude that when
the higher twist effects are taken into account and the S-D mixing
is considered as well, the branching ratio of $B\to \psi(3770)K$ can
become large enough to fit experimental data. If a smaller value for
$|t|$ is used, the calculated decay rates are somewhat smaller, but
still much more improved than the previous calculation. Here, the
soft scattering effects in the spectator interactions have played an
essential role. ¡¡

\section{Discussions and Summary}

In this paper, we study the $B^+\rightarrow\psi(3770)K^+$ decay
within the QCD factorization framework. If we treat $\psi(3770)$
as a pure $1^3D_1$ state and use the leading-twist approximation
for the kaon, we only get a very small branching ratio
${\mathrm{Br}} (B \rightarrow \psi^{''} K) =$$ 1.13 \times
10^{-5}$, which is about 40 times lower than the experimental
data.

We further introduce the $S-D$ mixing, combined with the calculation
for the $B^+\rightarrow\psi(3686)K^+$ decay, but still use the
leading-twist approximation for the kaon, then by fitting the
observed ratio of $B^+\rightarrow\psi(3770)K^+$ to
$B^+\rightarrow\psi(3686)K^+$, we find the required mixing angle to
be about $\theta=-26^\circ$ or $\theta=+59^\circ$. These mixing
angles are not consistent with that obtained from other experiments.
Moreover, the absolute branching ratio of
$B^+\rightarrow\psi(3770)K^+$ is still one order of magnitude lower
than the experimental data.

We then take the higher twist effects into account. By choosing
proper parameters to characterize the end-point singularities
related to the soft spectator interactions, and taking the S-D
mixing angle to be the widely accepted value $\theta=-12^{\circ}$,
we can get a much larger branching ratio, and it is then possible
to make the calculated rate of $B^+\rightarrow\psi(3770)K^+$ close
to the data.

We would like to emphasize that in the present calculation it is
the soft scattering effects in the spectator interactions that are
essential in enhancing the decay rates, though there exist
uncertainties for treating the soft singularities. Here, it might
be useful to discuss the possible connection between the inclusive
process $B\to\psi(3770)+ anything$ and the exclusive process
$B\to\psi(3770)K$. In fact, with the nonrelativistic QCD (NRQCD)
formulism\cite{bbl} it was pointed out\cite{yuanf} (see also
\cite{ko1}) that  for the D-wave charmonium inclusive production
in $B$ decays the color-octet $c\bar c$ operators in the weak
decay effective Hamiltonian may play the dominant role by
producing a color-octet $c\bar c$ pair at short distances, which
subsequently evolve to a color-singlet $c\bar c$ (the physical
charmonium) by emitting soft gluons at long distances. When the
emitted soft gluon interacts with and is absorbed by the spectator
light quark, the process becomes an exclusive one, such as
$B\to\psi(3770)K$ (the emitted soft gluons can of course hadronize
into light hadrons without interactions with the spectator quark).
If this picture makes sense, our observation in the present work
that the soft scattering effects in the spectator interactions
play the essential role in $B\to\psi(3770)K$ should be reasonable.

This may also be true for the $B$ exclusive decays involving
S-wave charmonia $J/\psi$\cite{cheng, chay} and
$\eta_c$\cite{song1}, where the calculations for $B\to J/\psi
(\eta_c) K$ without twist-3 soft spectator contributions are much
smaller than the observed rates, and the enhancement effect due to
higher twist is emphasized in\cite{cheng}. For the $B$ exclusive
decays involving P-wave charmonium states, the situation becomes
even more puzzling, that is, the measured nonfactorizable
$B\to\chi_{c0}K$ decay rate\cite{bellep0, babarp0} is large, about
an order of magnitude larger than that of another two
nonfactorizable decays $B\to\chi_{c2}K$\cite{babarp2} and $B\to
h_cK$\cite{bellehc}. These are not compatible with predictions
based on the final state rescattering model\cite{colangelo}. Some
of these decays are also studied in the PQCD approach with $k_t$
factorization\cite{li}, and in the light-cone sum rule
approach\cite{huang}. In QCD factorization it is found that for
the $B$ exclusive decays involving P-wave charmonium states, there
exist infrared divergences in the QCD vertex
corrections\cite{song2, song3}. However, if the twist-3 soft
spectator interactions dominate, we might provide a possible
explanation for the puzzle related to $B\to\chi_{c0}(\chi_{c2},
h_c) K$ decays, and this result will be presented
elsewhere\cite{meng}.

\section*{Acknowledgements}

We thank H.Y. Cheng for useful comments. This work was supported
in part by the National Natural Science Foundation of China (No
10421503), and the Key Grant Project of Chinese Ministry of
Education (No 305001).


\begin{thebibliography}{99}

\bibitem{bes}   J.Z. Bai et al.(BES Collaboration), Phys. Lett. B605,
63(2005); M. Ablikim et al. (BES Collaboration), hep-ex/0605105;
hep-ex/0605107.

\bibitem{cleo2} N. E. Adam et al. (CLEO Collaboration), Phys. Rev.
Lett. 96, 082004 (2006).

\bibitem{cleo1}  G.S. Adams et al. (CLEO Collaboration), Phys. Rev. D 73 (2006)
012002.

\bibitem{kuang} Y.P. Kuang, Phys. Rev. D65, 094024(2002); Front. Phys. China 1, 19(2006) (hep-ph/0601044).


\bibitem{cleo3}  T.E. Coan et al. (CLEO Collaboration), Phys. Rev. Lett. 96 (2006)
182002;  R.A. Briere et al. (CLEO Collaboration), hep-ex/0605070.





\bibitem{rosner} J.L.Rosner, Phys. Rev. D64, 094002 (2001).

\bibitem{yuan} C.Z.Yuan, hep-ex/0605078.

\bibitem{belled} K. Abe et al.(Belle Collaboration), Phys. Rev. Lett. 93 (2004) 051803.

\bibitem{pdg} S. Eidelman et al. [Particle Data Group], Phys. Lett. B 592 (2004)1.



\bibitem{ding} Y.B. Ding, D.H. Qin, and K.T. Chao, Phys. Rev. D 44, 3562 (1991).

\bibitem{eichten} E. Eichten et al., Phys. Rev. D21, 203 (1980); D17, 3090 (1978).

\bibitem{heikkila} K. Heikkila et al., Phys. Rev. D29, 110 (1984).













\bibitem{BSW}
M.~Bauer, B.~Stech and M. Wirbel, Z. Phy. C {\bf 34}, 103 (1987).

\bibitem{liu}
K.Y. Liu and K.T. Chao, Phys. Rev. D70, 094001 (2004).

\bibitem{BBNS1} M. Beneke, G. Buchalla, M. Neubert and
C.T. Sachrajda, Nucl. Phys. B  {\bf 591 } (2000)313.

\bibitem{BBNS2} M. Beneke, G. Buchalla, M. Neubert and
C.T. Sachrajda, Phys. Rev. Lett. {\bf 83 } (1999)1914.

\bibitem{BBNS3} M. Beneke, G. Buchalla, M. Neubert and
C.T. Sachrajda, Nucl. Phys. B  {\bf 606 } (2001)245.

\bibitem{BBL2}  G. Buchalla, A.J. Buras and M.E. Lautenbacher,
Rev. Mod. Phys. {\bf 68} (1996)1125.

\bibitem{kuhn} J.H. K\"{u}hn, Nucl. Phys. B  {\bf 157} (1979)125;
B. Guberina et al., Nucl. Phys. B  {\bf 174} (1980)317.

\bibitem{quig}  E.J. Eichten and C. Quigg, Phys. Rev. D {\bf 52} (1995)1726.

\bibitem{chay} J. Chay and C. Kim, hep-ph/0009244.

\bibitem{cheng}
H.Y.~Cheng and K.C.~Yang, Phys.\ Rev.\ D {\bf 63} (2001)074011.

\bibitem{ball} P. Ball, JHEP {\bf 9809} (1998)005.

\bibitem{gray}A. Gray et al., Phys. Rev. Lett. {\bf 95},
212001 (2005).

\bibitem{bbns}M. Beneke, G. Buchalla, M. Neubert and C. T. Sachrajda,
hep-ph/0007256; M. Beneke, J. Phys. {\bf G27} (2001) 1069.

\bibitem{bbl}  G.T. Bodwin, L. Braaten, and G. P. Lepage, Phys. Rev. D51, 1125 (1995).

\bibitem{yuanf} F. Yuan, C.F. Qiao, and K.T. Chao, Phys. Rev. D56, 329
(1997).

\bibitem{ko1} P.W. Ko, J. Lee, and H.S. Song, Phys. Lett. B395,
107(1997).


\bibitem{song1} Z. Song, C. Meng and K.T. Chao, Eur. Phys. J. C 36, 365 (2004).

\bibitem{bellep0} A. Gamash et al. (Belle Collaboration), Phys. Rev.
D71, 092003 (2005).

\bibitem{babarp0} B. Aubert et al. (BaBar Collaboration), Phys. Rev.
D69, 071103 (2004).

\bibitem{babarp2}  B. Aubert et al. (BaBar Collaboration), Phys.
Rev. Lett. 94, 171801 (2005).

\bibitem{bellehc}  F. Fang et al. (Belle Collaboration),
hep-ex/0605007.

\bibitem{colangelo} P. Colangelo, F. De Fazio, and T.N. Pham,
Phys. Lett. B542, 71 (2002); Phys. Rev. D 69, 054023 (2004).

\bibitem{li} C.H. Chen and H.n. Li, Phys. Rev. D71, 114008 (2005).

\bibitem{huang} B. Melic, Phys. Lett. B 591, 91 (2004); Z.G. Wang,
L. Li, and T. Huang, Phys. Rev. D70, 074006 (2004).



\bibitem{song2} Z. Song and K.T. Chao, Phys. Lett. B 568, 127
(2003).

\bibitem{song3} Z. Song, C. Meng, Y.J.Gao and K.T. Chao,
Phys. Rev. D {\bf 69}, 054009 (2004).

\bibitem{meng} C. Meng, Y.J. Gao, and K.T. Chao, hep-ph/0607221; hep-ph/0502240.




\end{thebibliography}
\end{document}